\documentstyle[sprocl,epsfig]{article}

\def\alb{\overline{\alpha}}
\def\be{\begin{equation}}
\def\ee{\end{equation}}
\def\bea{\begin{eqnarray}}
\def\eea{\end{eqnarray}}

\newcommand{\lsim}{\raisebox{-0.07cm}{$\,\stackrel{<}{{\scriptstyle
 \sim}}\, $} }

\newcommand\GeV{\,\mbox{GeV}}

\newcommand\om{\,\mbox{\boldmath $\omega$}}
\newcommand\ALM{\,\mbox{\boldmath $\alpha_s$}}

\begin{document}

\vspace*{-0.5cm}
\begin{flushleft}
DESY 98--036  \hfill {\tt hep-ph/9806368} \\
WUE-ITP-98-017\\
April~1998\\
\end{flushleft}
\vspace*{0.5cm}

\title{THE UNPOLARIZED GLUON ANOMALOUS DIMENSION AT SMALL 
{\boldmath $x\,$}\footnote{
Work supported in part by EU Network contract FMRX-CT98-0194 and by the 
German Federal Ministry of Research under contract No.\ 05 7WZ91P (0).\\
To appear in the Proceedings of the International Conference DIS'98,
Brussels, April 1998.}}

\author{J. BL\"UMLEIN, V. RAVINDRAN, W.L. VAN NEERVEN$\,$\footnote{On
leave of absence from Instituut-Lorentz, University
of Leiden, The Netherlands}}

\address{DESY-Zeuthen, Platanenallee, D-15735 Zeuthen, Germany
}

\author{A. VOGT}

\address{Institut f\"ur Theoretische Physik, Universit\"at W\"urzburg,
\\ Am Hubland, D-97074 W\"urzburg, Germany
}


\maketitle\abstracts{
We discuss the quantitative consequences of the resummation of the
small-$x$ contributions to the anomalous dimensions beyond 
next--to--leading order in $\alpha_s$ and up to next order in $\ln(1/x)$
(NL$x$) in a framework based on the renormalization group equations. 
We find large and negative effects leading to negative values for the 
{\sf total} splitting function $P_{gg}(x,\alpha_s)$ already for $x 
\lsim 0.01$ at $Q^2 \simeq 20 \GeV^2$. Terms less singular than those 
under consideration turn out to be quantitatively as important and 
need to be included. We derive the effects of the conformal part of the 
NL$x$ contributions to the anomalous dimensions and discuss the 
exponent $\omega$ describing the $s^{\omega}$ behavior of inclusive 
cross sections.}

\section{Introduction}

The scaling violations of deep--inelastic structure functions are
described by renormalization group equations (RGE's) stemming from
the ultraviolet
singularities of the corresponding local operators of a given twist.
For the twist-2 contributions, these RGE's are equivalent to the ones
derived from
 mass factorization. In the following we will consider only
these twist-2 terms, and study the impact of their resummed small-$x$
contributions on the evolution of structure functions. The
non-perturbative input distributions at a scale $Q^2_0$ {\sf factorize}.
Furthermore, both the coefficient functions $c_{k,l}(x,Q^2/\mu^2)$ and
the splitting functions $P_{ij}(x,\mu^2)$ are completely known up to
two--loop order. Therefore they can be accounted for in {\sf complete}
form, whereas the resummed small-$x$ terms are considered beyond
this order in addition.

The generating functional for the leading small-$x$ order (L$x$) has
been known for a long time~\cite{BFKL}. More recently the resummed
NL$x$ coefficient functions and quarkonic anomalous dimensions were
calculated~\cite{CH}. Their structure and their numerical impact on the
evolution of the deep--inelastic structure functions, as well as the
consequences of the NL$x$ anomalous dimension $P_{gg}$, have been
discussed in great detail, cf.~ref.~\cite{BV97}. As the derivation of
$P_{gg}$ at NL$x$ order~\cite{CC1} was completed recently~\cite{FL,CC2},
the present paper provides an update of previous results~\cite{BV97},
the quantitative changes with respect to that analysis being rather
minor. We first describe the structure of $P_{gg}$ as emerging from the
NL$x$ resummation, then present some numerical results, and finally
discuss consequences on the intercept $\omega(\alpha_s)$ describing the
$s^{\,\omega}$ behavior of inclusive cross sections, which sometimes is
qualitatively interpreted also as the `rising' power of structure
functions at small $x$. As already pointed out several years ago~\cite
{SUB1}, and in various more recent numerical studies of a large variety
of small-$x$ resummations~\cite{SUB2,BV97}, we find that, also in the
present case, meaningful results can only be obtained including
the less singular orders which turn out to be {\sf quantitatively as
important} as the known ones.

\section{The structure of $P_{gg}^{{\,\rm NL}x}$}

In ref.~\cite{FL} a Bethe--Salpeter equation with an infrared finite
kernel at $O(\alpha_s^2)$ was derived summing the leading pole $\propto
1/(N-1)$ contributions. Its diagonalization can be performed a by Mellin
transform in the transverse--momentum space,
\begin{eqnarray}
\int \! d q_2^2 \, (q_2^2/q_1^2)^{\gamma-1}  K(q_1,q_2) =
\frac{N_c \alpha_s(q_1^2)}{\pi} \left[ \chi_0(\gamma)
+ \frac{N_c \alpha_s(q_1^2)}{\pi} \delta(\gamma) \right]~.
\label{eig}
\end{eqnarray}
The r.h.s.\ of eq.~(\ref{eig}) can be related to the resummed anomalous
dimension $\gamma_{gg}(N,\alpha_s)$ in NL$x$ order, see, e.g., ref.\
 \cite{BV97}. This can be easily seen in the {\sf conformal limit}
\cite{CL} $m = 0$, $\beta(\alpha_s) = 0$, see ref.~\cite{BRN1} for 
details, where the coupling constant $\alpha_s$ is a fixed number. In 
L$x$ order the Bethe--Salpeter equation~\cite{BFKL} obeys this 
criterion. Therefore the solution $\gamma$ of the resulting eigenvalue 
equation,
\begin{eqnarray}
1 = \frac{\alb_s}{N-1} \chi_0(\gamma),~~~\chi_0(\gamma) = 2 \psi(1)
- \psi(\gamma) - \psi(1- \gamma),~~\alb_s = \frac{N_c}{\pi} \alpha_s~,
\label{eqLXE}
\end{eqnarray}
represents the corresponding part of the anomalous dimension at all
orders of the coupling constant.

In NL$x$ order the situation is more complicated. Instead of
eq.~(\ref{eqLXE}) one obtains
\begin{eqnarray}
1 = \frac{\alb_s}{N-1} \left [\chi_0(\gamma_+) - \frac{\alpha_s}{4}
\delta(\gamma_+,q_1^2,\mu^2) \right]~,
\label{eqNLXE}
\end{eqnarray}
with $\gamma_+$ the larger eigenvalue of the singlet anomalous
dimension matrix and
\begin{eqnarray}
\delta(\gamma,q_1^2,\mu^2) = \frac{\beta_0}{3} \chi_0(\gamma) \ln
\left(\frac{q_1^2}{\mu^2}\right)
\! + \!\left[\frac{\beta_0}{6} \! + \!\frac{d}{d\gamma} \right]\!
\left[\chi_0^2(\gamma) + \chi'_0(\gamma) \right] + \chi_1^{\rm symm}
(\gamma)\, .
\label{break}
\end{eqnarray}
All terms but the last one are related to the breaking of conformal
invariance in NL$x$ order, cf.\ also ref.~\cite{CC2}. The second term 
changes for a different scale choice. Thus {\sf only} $\chi_1^{\rm 
symm}$ can dealt with as in the L$x$ case. Note that the NL$x$ quarkonic
anomalous dimensions~\cite{CH} result from a conformally invariant 
kernel in the $Q_0$-scheme~\cite{Q0}, they are therefore free of this 
problem.

The construction used to relate the solution of eq.~(\ref{eqNLXE}) to 
the NL$x$ anomalous dimension beyond the conformally invariant part
is~\cite{CC3} to absorb the term $\propto \ln (q_1^2/\mu^2)$ into the
coupling constant and to keep all the other terms. The implicit equation
(\ref{eqNLXE}) is then solved representing $\gamma$ as an infinite
series in $\alpha_s/(N-1)$ for complex values of $N$. All details for
this solution as well a discussion of the mixing problem, which has to 
be solved order by order in $\alpha_s$ in the singlet case, have been 
given in ref.~\cite{BV97}. Here we would like to add only a few remarks 
on the various contributions~\cite{FL} to
\begin{eqnarray}
\chi_1(\gamma) &=& \delta(\gamma,q_1^2,\mu^2) - \frac{\beta_0}{3}
\chi_0(\gamma) \ln\left(\frac{q_1^2}{\mu^2}\right) =
\chi_1^{(1)}(\gamma)  +  \chi_1^{(2)}(\gamma) +  \chi_1^{(3)}(\gamma),
\nonumber\\
\chi_1^{(1)}(\gamma) &=&  \frac{\pi^2}{\sin^2(\pi \gamma)}\frac{
\cos(\pi\gamma)}{1 - 2 \gamma} (22 -\beta_0),\nonumber\\
\chi_1^{(2)}(\gamma) &=&  \frac{\pi^2}{\sin^2(\pi \gamma)}\frac{
\cos(\pi\gamma)}{1 - 2\gamma} \left[ \frac{\gamma(1-\gamma)}{(1+2\gamma)
(3-2\gamma)}\left(1 + \frac{N_f}{3}\right)\right]
\nonumber\\
&-&
\left(\frac{67}{9} - 2\zeta(2) -\frac{10}{27} N_f\right) \chi_0(\gamma)
+ 4\Phi(\gamma) - \frac{\pi^3}{\sin^2(\pi\gamma)},\nonumber\\
\chi_1^{(3)}(\gamma) &=&  \left [\frac{\beta_0}{6} + \frac{d}{d\gamma}
\left[\chi_0^2(\gamma) +\chi_0'(\gamma) \right]\right] - 6 \zeta_3~.
\end{eqnarray}
The terms $\chi_1^{(i)}(\gamma)$ contribute for the first time in
$i$-loop order. Note that the $\beta_0$-term $\chi_1^{(1)}$ is due to
the gluon self--energy, and has nothing to do with running coupling
effects, unlike the one in $\chi_1^{(3)}$. Therefore in the first two
orders in $\alpha_s$ only conformally
invariant terms contribute to the gluon
anomalous dimension in the small-$x$ limit even in NL$x$ order. These 
terms are unique in the class of DIS schemes. Let us note that the 
function $\Phi(\gamma)$ obeys the representation
\begin{eqnarray}
\Phi(\gamma) = \frac{1}{\gamma} \sum_{l=2}^{\infty} (-1)^l \zeta(l)
\gamma^{l-2} + \sum_{k=0}^{\infty}\left[\frac{2\pi^2}{3}\eta(2k+2)
+c_{2k+1}\right]\gamma^{2k+1}
\end{eqnarray}
with $\eta(k)=\zeta(k)\left[1-2^{1-k}\right]$ and $c_k = -2/k!\int_0^1 
dz \ln^k(1/z) {\rm Li}_2(z)/(1+z)$, which are new transcendentals
\cite{BK97}. At 3-loop order the running coupling effects contribute 
for the first time. It should be mentioned that concerning the 
$\zeta(3)$-term differing results exist in the literature~\cite{FFK,KK}.
The gluon Regge-trajectory was recently recalculated \cite{BRN1} and 
agreement with the results of ref.~\cite{FFK} was found.
The numerical effect on both the anomalous dimension $\gamma_{gg}$ in 
3-loop order and the intercept $\omega$ arising from the difference of 
the results~\cite{FFK,KK} is not negligible, see ref.~\cite{BRN1}. 
Finally, starting with 4-loop order, scheme--specific terms by which,
e.g., the $Q_0$ and the usual DIS scheme differ, and other effects due 
to the running of $\alpha_s$, which are not discussed in ref.~\cite{FL},
contribute to the anomalous dimension (for details see ref.~\cite
{BV97}). The latter effects are numerical very large.
\begin{minipage}[b]{11.5cm}
\vspace*{1mm} 
\parbox[c]{7cm}{
\mbox{\epsfig{file=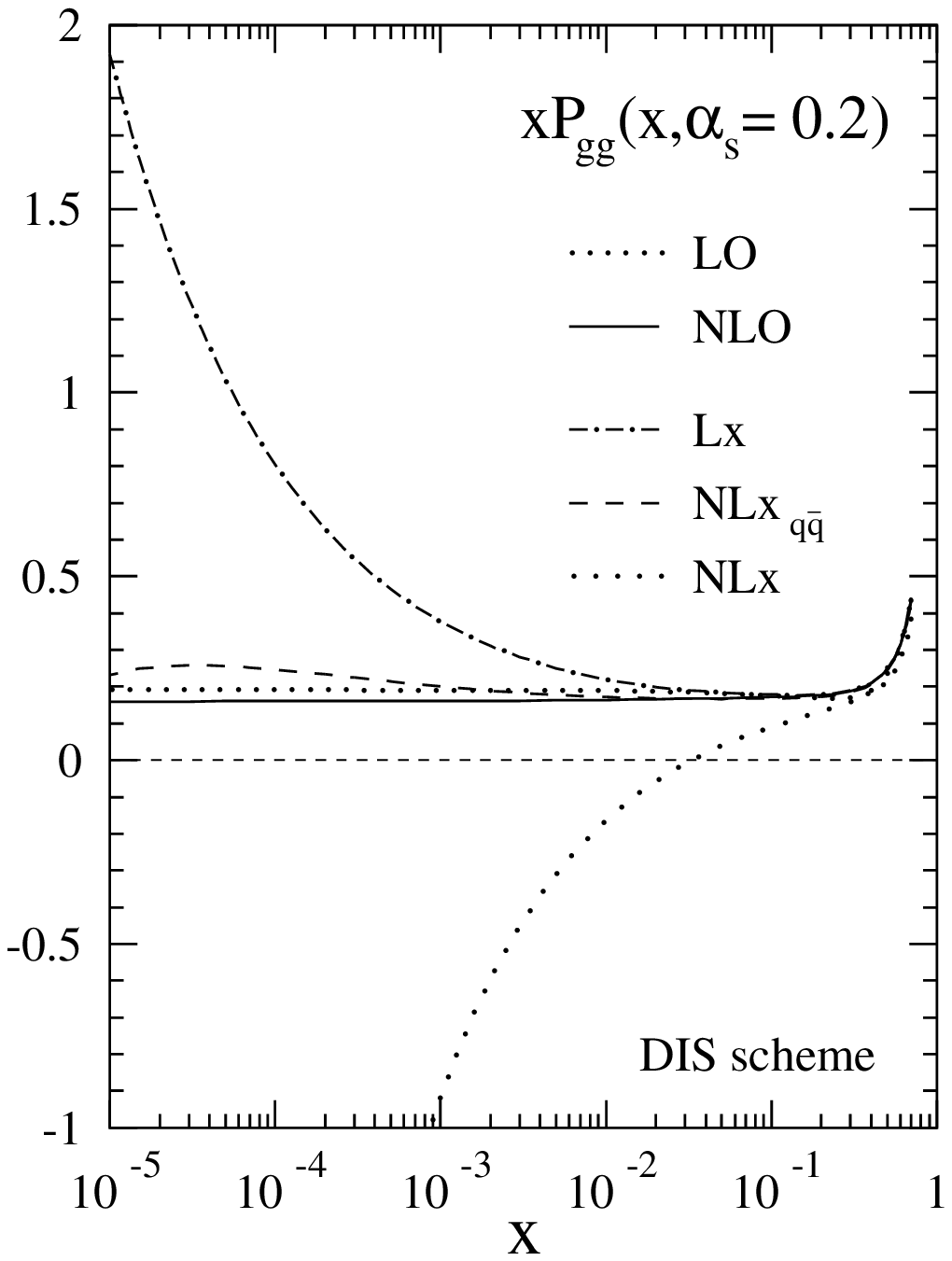,width=7cm,angle=0}
}}~~
\parbox[c]{4cm}{
\begin{center}
\begin{tabular}{||r|r||}
\hline\hline
 & \\[-4mm]
\multicolumn{1}{||c|}{$k$} &
\multicolumn{1}{c||}{$\Delta g_{k,gg}^{\, gg}$} \\
 & \\[-4mm] \hline\hline  & \\[-4mm]
  0 &--1.65000$\,$E+1\\
  1 &  0.00000$\,$E+0\\
  2 &--2.78734$\,$E+1\\
  3 &--2.25279$\,$E+2\\
  4 &--1.65583$\,$E+2\\
  5 &--7.24788$\,$E+2\\
  6 &--3.14501$\,$E+3\\
  7 &--3.49585$\,$E+3\\
  8 &--1.51028$\,$E+4\\
  9 &--4.91970$\,$E+4\\
 10 &--7.46877$\,$E+4\\
 11 &--2.99245$\,$E+5\\
 12 &--8.31843$\,$E+5\\
 13 &--1.59528$\,$E+6\\
 14 &--5.82155$\,$E+6\\
 15 &--1.49497$\,$E+7\\
 16 &--3.37088$\,$E+7\\
 17 &--1.12828$\,$E+8\\
 18 &--2.81522$\,$E+8\\
 19 &--7.03719$\,$E+8\\[0.5mm]
\hline\hline
\end{tabular}

\end{center}
}
\\ \mbox{}
\end{minipage}

\noindent
{\small\sf Figure~1: The $x$-dependence of the the splitting function
$P_{gg}(x,\alpha_s)$ at a typical value of $\alpha_s$. The cumulative 
effect of the different orders is shown by consecutively adding the LO, 
NLO, L$x$, ${\rm NL}x_ {q\overline{q}}$, and NL$x$ contributions.}  

\vspace*{2mm} 
In the table above an update is given of the coefficients $\Delta 
g_{k,gg}^{\, gg}$ of table~1 in ref.~\cite{BV97} is given, based on the 
recent results in refs.~\cite{FL,CC2}. The various contributions to the 
splitting function $P_{gg}(x,\alpha_s)$ are illustrated for the 
kinematic range at HERA in Figure~1. The NL$x$ contributions are very 
large, leading to negative values for $P_{gg}(x,\alpha_s)$ for $x 
\simeq 0.01$ and $Q^2 \simeq 20 \GeV^2$, thus destroying the 
probabilistic interpretation of $P_{gg}(x)$ in LO. As demonstrated in 
ref.~\cite{BV97}, yet unknown less singular terms are expected to 
appear at the same size but with different sign. They are therefore 
quantitatively as important for the anomalous dimension in the small-$x$
range as the L$x$ and NL$x$ terms.

\section[xxx]{The `rising' power \om(\ALM)}

The leading $s \simeq \ln(1/x)$ behavior of the scattering cross section
can be described by
\begin{eqnarray}
\sigma(s) \propto s^{\omega(\alpha_s)}~.
\end{eqnarray}
In general the $s$-dependence of the cross section is {\sf not only}
due to perturbative contributions but also determined by the impact
factors. If one neglects non-perturbative effects and discusses, in an 
{\sf informal way}, only $\omega(\alpha_s)$ as a perturbative, idealized
value, this exponent is obtained evaluating $\chi(\gamma)$ for the 
branch point $\gamma = 1/2$~\cite{FL}, for $N_f =4$ resulting 
in$\,$\footnote{ Note a numerical error in eq.~(16) of ref.~\cite{FL}.}
\begin{eqnarray}
\omega(\alpha_s) = 2.65 \,\alpha_s \left[ 1 - 6.36 \,\alpha_s \right]~.
\label{eqom}
\end{eqnarray}
The quadratic relation~(\ref{eqom}) leads to a maximum $\omega_{\rm max}
= 0.10$ at $Q^2\simeq 8.7\times 10^{\, 6}\GeV^2$ which is comparable to
the phenomenological value $\omega_{\,\rm DL} = 0.0808$ for the {\sf soft}
pomeron~\cite{DOLA}. Moreover, in most of the small-$x$ kinematic range 
at HERA it takes negative values (for $Q^2 \lsim 600 \GeV^2)$, e.g., 
$\omega(20 \GeV^2) \simeq -0.35$, while the leading order value amounts 
to $\omega(20 \GeV^2) \simeq +0.64$. Since second the order correction 
yield such a drastic modification of $\omega$, yet unknown less singular
terms are to be expected to change this result significantly once again.

In the above exponentiation, both the conformally invariant parts
in $\chi(\gamma)$ and those which are due to the breaking terms of
conformal invariance
have been included. The latter contributions are related to the
terms $\left [\chi^2_0(\gamma) + \chi'_0(\gamma)\right]$, cf.\ 
eq.~(\ref{break}). It is known that the conformally invariant terms
exponentiate~\cite{BRN1}, however, the effect of the latter terms is 
less clear. If, as an example, one completely
discards all terms which are due to the breaking of conformal invariance
(see above) one
obtains, (again for $N_f=4$)
\begin{eqnarray}
\omega_{\rm conf}(\alpha_s) &=& 2.65 \,\alpha_s \left[ 1 - 2.55 
\,\alpha_s \right]
\label{eqomc}
\end{eqnarray}
with a maximum value of $\omega_{\rm conf}^{\rm max} = 0.26$ at 
$Q^2_{\rm max} \simeq 90 \GeV^2$. Furthermore $\omega_{\rm conf}$
is positive for $Q^2 \simeq 2 \GeV^2$. Note that both values above are
{\sf idealized} ones, and clearly the effects of the less singular 
terms, those due to the running of $\alpha_s$, and the non-perturbative 
input do need much further understanding for this quantity.

\vspace*{-1mm}

\end{document}